\documentstyle[aps,epsf,rotate,multicol]{revtex}

\begin{document}

\draft

\title{Comment on: Kinetic Roughening in Slow Combustion of Paper }

\author{ Lu\'{\i}s~A.~Nunes Amaral$^{(1)}$ and Hern\'an A. Makse$^{(2)}$}

\address{
$^{(1)}$Condensed Matter Theory, Physics Dept., Massachusetts Institute
 of Technology, Cambridge, MA 02139 \\ 
$^{(2)}$Schlumberger-Doll Research, Old Quarry Road, Ridgefield, CT 06877}

\date{Phys. Rev. Lett. {\bf 80}, 5706 (1998)}

\maketitle


\begin{multicols}{2}

In a recent Letter, Maunuksela {\it et al.\/} present experiments on
the combustion of paper \cite{Maunuksela}.  They observe the
roughening of the burning front and study the scaling properties of
the correlation function of the height of the burning front.
Maunuksela {\it et al.\/} find good agreement between the measured
exponents characterizing the front roughening and the predictions of
the Kardar-Parisi-Zhang (KPZ) equation \cite{KPZ86}, $\chi_{KPZ} =
1/2$ ($\chi$ is the roughness exponent which characterizes the scaling
of the saturated height-height correlation function $G(r)$ with
distance $G(r) \sim r^{2\chi}$).  Reference~\cite{Maunuksela} also
comments on the results of earlier experiments on paper burning by
Zhang {\it et al.\/} \cite{Zhang} which measured $\chi \approx 0.71$,
but offers no explanation for the difference in the measured value of
the exponent.

Here, we show that the results of Maunuksela {\it et al.\/}
\cite{Maunuksela} and of Zhang {\it et al.\/} \cite{Zhang} may be {\it
both} understood under the framework of interface motion in disordered
media \cite{DPD,Amaral,AmaralPRE,Laszlo}.

For many experimental cases, the dominant source of noise in the
dynamics is the {\it disorder in the medium\/} which is not time
dependent.  The universality classes for interface motion in
disordered media have been identified and the values of the exponents
are known, especially for $(1+1)$ dimensions
\cite{DPD,Amaral,AmaralPRE,Laszlo}.  One of the universality classes
for interface motion in disordered media \cite{Laszlo} can be
described by the directed percolation depinning (DPD) model
\cite{DPD,Amaral,AmaralPRE}. For a driving force $F$ smaller than a
critical value $F_c$, which defines the depinning transition, the
interface eventually stops, and the scaling of $G(r)$ is characterized
by an exponent $\chi_{DPD} \approx 0.63$ \cite{DPD}.  If, on the other
hand, $F > F_c$, the interface moves with an average velocity $v \sim
(F-F_c)^\theta$ (where $\theta$ is a critical exponent), and $G(r)$
presents two scaling regimes in the steady-state.  For length-scales
{\bf smaller} than the correlation length given by the disorder of the
medium $\xi\sim (F-F_c)^{-\nu_\parallel}$ (where $\nu_\parallel$ is a
critical exponent), $G(r)$ scales with an exponent $\chi'_{DPD}
\approx 0.75$, while for length-scales {\bf larger} than $\xi$, $G(r)$
scales with the KPZ exponent, $\chi_{KPZ} = 1/2$ \cite{AmaralPRE}.

\begin{figure}
 \centerline { \vbox{ \hbox
     {\epsfxsize=6.7cm \epsfbox{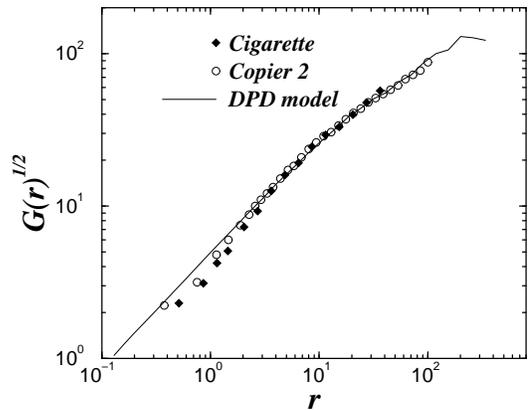} } }
     }
\narrowtext
\caption{ Scaling of the height-height correlation function for the
        DPD model \protect\cite{AmaralPRE} and the data from the paper
        burning experiments of Ref.~\protect\cite{Maunuksela} for
        cigarette and copier papers.  The results from the model were
        translated so as to make the crossover positions coincide.  It
        is visually apparent that the two curves have the same two
        scaling regimes.}
\label{f-data}
\end{figure}

To check that the results reported by Maunuksela {\it et al.\/} are
consistent with the above theory, we digitize the data reported in
\cite{Maunuksela} and plot it in Fig.~\ref{f-data} along with the predictions
of the DPD model (see Refs.~\cite{DPD,Amaral,AmaralPRE}).  It is visually
apparent that the experimental and theoretical data sets have the same two
scaling regimes as discussed in Ref.~\cite{AmaralPRE}.  Furthermore, the same
theory may explain why Zhang {\it et al.}  found no crossover to a KPZ
dominated regime.  Zhang's experiments were performed near the depinning
transition, which implies that $\xi$ was nearly as large as the system size.
Thus, the quenched disorder dominates the scaling of $G(r)$ leading to the
observation of only one scaling regime with a roughness exponent $\chi
\approx \chi'_{DPD}$.  On the other hand, in the experiments reported by
Maunuksela {\it et al.}, a shorter correlation length $\xi$ is found, so that
quenched disorder dominates only for small length scales where an exponent
with values close to $\chi'_{DPD}$ was measured.  For length scales larger
than $\xi$, time-dependent disorder dominates, and the KPZ exponent
$\chi_{KPZ}$ was measured, in agreement with the theory
\cite{AmaralPRE}.  These results suggest that the DPD universality class
may provide a compelling explanation of the experimental findings of
Refs.~\cite{Maunuksela,Zhang}.

\end{multicols}

\end{document}